\documentclass[journal]{IEEEtran}
\usepackage[utf8]{inputenc}
\usepackage{graphicx}
\usepackage{gensymb}
\usepackage{listings}
\usepackage{caption}
\usepackage{subcaption}
\usepackage{cite}
\usepackage{algpseudocode}
\usepackage{algorithm}
\usepackage[section]{placeins}
\usepackage{url}
\usepackage{authblk}

\newcommand{\RomanNumeralCaps}[1]
    {\MakeUppercase{\romannumeral #1}}
    
\author{Aparajithan~Nathamuni-Venkatesan, Ram-Venkat~Narayanan, Kishore~Pula, \\ Sundarakumar~Muthukumaran and Ranga~Vemuri}
\affil{Digital Design Environments Lab, ECE Department\\University of Cincinnati, Cincinnati, Ohio, USA\\nathaman@mail.uc.edu, narayart@mail.uc.edu, pulake@mail.uc.edu, \\ 
muthuksr@mail.uc.edu, vemurir@mail.uc.edu}
\title{Word-Level Structure Identification In FPGA Designs Using Cell Proximity Information}

\algrenewcommand\algorithmicindent{0.8em}
\begin{document}
\pagenumbering{gobble} 

\maketitle
\begin{abstract}
    Reverse engineering of FPGA based designs from the flattened LUT level netlist to high level RTL helps in verification of the design or in understanding legacy designs. We focus on flattened netlists for FPGA devices from Xilinx 7 series and Zynq 7000. We propose a design element grouping algorithm that makes use of the location information of the elements on the physical device after place and route. The proposed grouping algorithm gives clusters with average NMI of 0.73 for groupings including all element types. The benchmarks chosen include a range of designs from communication, arithmetic units, processors and DSP processing units.
\end{abstract}
\section{Introduction}

\IEEEPARstart{T}{h}ere are several applications of reverse engineering a design from the device level to the RT level. One may want to understand legacy designs for which the source code is unavailable. There is also a need for reverse engineering FPGA designs in order to detect any malicious logic such as hardware trojans inserted in the design. Trojans can be inserted in the FPGA supply chain during specification and design, FPGA tool flow or during distribution \cite{bhunia2018hardware}.

Reverse engineering of FPGAs starts with reading out the bitstream used to program the FPGA. The LUTs and design elements are mapped onto the bitstream and the netlist is extracted. The flattened netlist is then analyzed to understand the high level functionality of the design. The process of understanding high level functionality of a netlist is called specification discovery \cite{azriel2021survey}. 

In specification discovery, the state and data registers are identified first along with identifying module boundaries and partitioning the netlist. The controllers are separated from the netlist to extract the FSM function. The data registers and word level structures are also identified in the netlist and the accompanying datapath blocks are recognized. The netlist would also be partitioned using clustering algorithms to compare with modules from a reference library. The matching step can make use of graph similarity, approximate matching of input and output ports of a block in a block diagram or matching the Boolean functions of the sub-circuits \cite{azriel2021survey, cakir2019revealing}.

The difficulty of performing specification discovery arises because of information lost or hidden in the flattened netlist. The synthesis tools perform various optimization techniques on the original design. Functional blocks may overlap among modules or the original function could be optimized to remove redundancies. In case of FPGAs, LUTs absorb considerable amount of logic into a single LUT which could cause structural information to be nearly lost.

We present the first work on design element grouping in fully flattened netlists for FPGAs which uses location information of the design elements utilized on the FPGA device. The algorithm presented makes use of proximity information and groups nodes at the site level and the word level. The algorithm also includes the graph edit distance metric \cite{fyrbiak2019graph} to ensure the structural similarity of the nodes and supernodes in the groupings. Individual bits in word level structures are expected to be operated by similar if not exactly the same logic having similar if not exactly the same structures.

Section \RomanNumeralCaps{2} discusses the required background behind the work. Section \RomanNumeralCaps{3} reveals the details of the proposed algorithm and Section \RomanNumeralCaps{4} shows the experimentation and the results obtained. Section \RomanNumeralCaps{5} concludes the discussion on the proposed algorithm.
\section{Relevant Work}

Fyrbiak et al. \cite{fyrbiak2019graph} demonstrated that graph similarity is helpful in detecting hardware trojans in ASIC and FPGA netlists. Two graphs, if dissimilar, would likely differ in the functionality of the circuit. They state that exact graph matching would not work because of  optimization operations done on the netlist during logic synthesis. Errors could come from the process of netlist extraction as well. We make use of this graph similarity measure in our algorithm  by computing and using the graph edit distance between two subcircuits in the netlist.

Li et al. \cite{li2013wordrev} discuss an algorithm to group nets into words in an ASIC netlist. They first use a shape hashing technique to find candidate subcircuits that could belong to a word. This is done in order to reduce the number of computations required to group signals into words. They then find k-feasible cuts which look at the functionality of the logic cone. Here "k" is the number of inputs allowed in the transitive fan-in cone rooted at the target gate. After candidate words are found, a check is performed to see if the words can propagate forward or backward using symbolic evaluation. They state that few control wires should be nearby that would allow the target word to take the complemented or uncomplemented form of every bit in the source word. They apply their algorithm to combinational circuits and perform symbolic evaluation described by Bryant and Randal \cite{bryant1990symbolic} in topological order. Next, they examine the logic bound between words by using 2QBF formulation. They assume the input correspondence between the unknown module and the reference module. They find the logic values to be applied to the control wires such that the unknown module and the reference module are equivalent.

Meade et al. \cite{meade2018old} propose three algorithms, REBUS, REWIND and REPCA to group nets into words in ASIC designs. A seed is needed for the algorithm to run. Word propagation can happen backwards from the outputs. REWIND works like REBUS but groups nets based on their similarity scores. The similarity scores are given by examining the logic cone driving the net. REPCA uses certain properties of the nets, called princinpal components, examined to group nets into words. The properties can be neighboring gate type, the distance of a neighboring gate from the net or the fan-in and fan-out sizes of the neighboring gate. REPCA was found to work well on larger netlists and REBUS works well on all netlists. The algorithms are run with different parameters by the user to find the optimum result for a given design. They also recommend Normalized Mutual Information (NMI) as a good metric to measure the quality of clusters. We follow the same NMI metric to evaluate groupings in our work.

Werner et al. \cite{werner2018reverse} discuss an approach by which an ASIC netlist can be partitioned into their respective modules. The netlist is converted into a graph with weighted edges. The edge weights depend on the spatial distance of cells in the layout and graph centrality measures. In an ASIC design the standard cells are placed in close proximity to each other. The local wires would have higher edge weights and the global wires would have lower edge weights. Additionally, the type of standard cell and signal type is noted. They implement the Louvian method of graph partitioning discussed by Blondel et al. \cite{blondel2008fast}, which gives the user control over the number and the sizes of clusters expected. The reference block diagram and the resultant clusters obtained are compared. The comparison tells us which module a particular cluster belongs to.

Shi et al. \cite{shi2012extracting} illustrate a method to extract functional modules from flattened gate level netlists in ASIC designs. They simplify the netlist by removing all the state machines. They describe their methods starting from this point. They group registers into bundles based upon the control signals connected to the registers. The LSB bits are then mapped to the corresponding output bits based on a rule. Their rule states that a bit in a word does not depend on the bits in the higher positions in the word. They identify multiplexors and then use the register enable and multiplexor select lines to group more signals. They then use Reduced and Ordered Binary Decision Diagrams (ROBDDs) to check for equivalence between the unknown subcircuit and the library module.

Albartus et al. describe DANA \cite{albartus2020dana}, a tool which helps in grouping registers independent of the technology used. They work with the flip-flop dependency graph and describe ways to group registers based upon the neighboring node coloring. They have a preprocessing stage which gives an initial set of register groups. There is a processing stage where the registers are grouped depending on shared neighbor groups. Register groups are broken down if the neighbor groups are not common. They show almost correct grouping of registers at the word level. The authors state that in ASIC layouts, the registers are placed close to each other. The location data would tell us that these registers belong to the same word. The authors also state that in FPGAs, the registers have to be placed depending on the resource constraints which would make the location data would be unreliable.

There is a need for an algorithm that does not require the user to tune the parameters of the algorithm for every netlist. A set of parameters chosen should work well for all netlists and the user may not have full knowledge of the netlist beforehand in all cases. There is also a need for a grouping algorithm for elements other than registers in a flattened LUT level FPGA netlist.

\section{Procedure to obtain higher level groupings}

The main hypothesis behind the algorithm is that the design elements placed closer to each other on the FPGA device are closely related in functionality or hierarchy of the design. Design elements placed far apart have little in common. The location information that the algorithm works with is the absolute location of the element or cell on the device. The location data includes the site, tile and clock region of the location in which the element is placed in. 

\begin{algorithm}
\caption{Algorithm to group elements to obtain site level groupings}
\begin{algorithmic}
\footnotesize
\State $site\_groupings \gets \phi$ \Comment{Site level groupings}
\State $i \gets 1$
\While{$i \leq number\_of\_cells$}
    \State $cell \gets list\_of\_cells[i]$
    \If{$site\_groupings$ is empty}
        \State $N \gets \phi$ \Comment{Create a new grouping}
        \State $N \gets cell$ \Comment{Add cell to new grouping}
        \State $site\_groupings \gets N$ \Comment{Append to groupings}
    \Else
        \State $j \gets 1$
        \While{$j \leq number\_of\_site\_level\_groupings$}
            \State $N \gets site\_groupings[j]$
            \If{location of cell matches location of $N$}
                \State $N \gets cell$ \Comment{Add cell to site level grouping}
                \State \textbf{break}
            \Else
                \State $j \gets j+1$
            \EndIf
        \EndWhile
        \If{no site level grouping matched}
            \State $N \gets \phi$ 
            \State $N \gets cell$ 
            \State $site\_groupings \gets N$ 
        \EndIf
    \EndIf
    \State $i \gets i+1$
\EndWhile
\end{algorithmic}
\end{algorithm}

We illustrate the steps followed in our algorithm. Firstly, the elements belonging to the same site are grouped together. Let us label the elements with $n_{i}$, where i ranges from 1 to the number of elements utilized. Let us refer to these groupings by site with $N_{j}$, where j ranges from 1 to the number of sites utilized on the device. Next these groupings by site are grouped depending on the structural similarity of the subgraph and the proximity of the sites. Let us call these higher groupings by $G_{k}$, where k ranges from 1 to the number of higher level groupings obtained.

\begin{algorithm}
\caption{Algorithm to obtain higher level groupings}
\begin{algorithmic}
\footnotesize
\State $H \gets \phi$ \Comment{Higher level groupings}
\State $i \gets 1$
\While{$i \leq number\_of\_sites\_utilized$}
    \State $j \gets 1$
    \State $N_i \gets site\_groupings[i]$
    \If{$H$ is empty}
        \State $G \gets \phi$
        \State $G \gets N_i$
        \State $H \gets G$ \Comment{Add to higher level groupings}
    \Else
        \While{$j \leq$ length of $H$}
            \State $G \gets H[j]$
            \State $k \gets 1$
            \While{$k \leq$ number of groups in $G$}
                \State $N_k \gets G[k]$
                \State Calculate spatial distance between $N_i$ and $N_k$
                \State{Calculate edit distance between subgraphs contained in $N_i$ and $N_k$}
                \State Update the total spatial distance
                \State Update the total edit distance
                \State $k \gets k+1$
            \EndWhile
            \State $spatial\_dist\_avg \gets $ average spatial distance
            \State $edit\_distance\_avg \gets $ average graph edit distance
            \If{($spatial\_dist\_avg \leq spatial\_distance\_threshold$) $\textbf{and}$ ($edit\_dist\_avg \leq edit\_distance\_threshold$)}
                \State $G \gets N_i$ \Comment{Append to higher level grouping}
                \State \textbf{break}
            \Else
                \State $j \gets j+1$
            \EndIf 
        \EndWhile
        \If{no higher level grouping matched}
            \State $G \gets \phi$
            \State $G \gets N_i$
            \State $H \gets G$
        \EndIf
    \EndIf
    \State $i \gets i+1$
\EndWhile
\end{algorithmic}
\end{algorithm}

The distance measure used is the Manhattan distance. A weight is assigned to each site, tile and clock region with the relationship $w_{site} < w_{tile} < w_{clk\_region}$. The weight associated with a difference in site location is $w_{site}$. The weight given to the difference in the tile location is $w_{tile}$ and the weight given to difference in clock region is $w_{clk\_region}$. The relationship defined for the weights follows from the stated hypothesis about the relationship between elements based on the proximity. Elements placed in different clock region are likely to belong to different word level groups and $w_{clk\_region}$ is given the highest weight. Elements placed in different but neighboring tiles could belong to the same group and elements placed in different but neighboring sites could be in the same group. We have assigned $w_{site}=1$, $w_{tile}=5$ and $w_{clk\_region}=25$. A difference in site coordinate has a multiplier of $1$, a difference in tile coordinate has a multiplier of $5$ and a difference in clock region coordinate has a multiplier of $25$. We take the product of the weight and the Manhattan distance for each site, tile or clock region information. We add all the distances to get a total distance measurement that includes information at site, tile and clock region levels in the device hierarchy. We compute the total distance between the unassigned grouping by site $N\_{unassgn}$ and each grouping by site in the higher level grouping we currently are examining. We then take the average of the total spatial distance to compare against the set constraint on proximity. The threshold set for the average spatial distance is 100 units to accommodate larger word lengths. A more tighter constraint on the spatial distance allowed would break down the word into smaller sizes which would then not truly reflect the actual hierarchy in the design.

In addition to the proximity information, we utilize the graph similarity metric \cite{fyrbiak2019graph}, to allow only groupings based on site with similar graph structures to form the higher level groupings. We impose strict constraints on the graph edit distance. The average edit distance allowed is $3$ as the LUTs and CARRY elements can absorb considerable amount of logic into one node. We consider the set of nodes in a site and the subgraph containing those nodes in the netlist. We have a grouping $N_{assgn}$ that is already assigned a higher level grouping labeled $G_{k}$. We have an unassigned grouping $N_{unassgn}$ and compare with all the assigned groupings $N_{assgn}$s in $G_{k}$. We compute the average location distance and the average graph edit distances. The average location distance is first checked with the threshold set at $100\ units$ and the average edit distance is checked with the threshold of $3\ units$. When these criteria are met the unassigned grouping $N_{unassgn}$ is assigned to $G_{k}$, otherwise we examine the next grouping $G_{k+1}$. If no higher level groupings match, we create a new higher level group and add the unassigned grouping $N_{assgn}$ to this higher level grouping.

The weights for site, tile and clock region are $1$, $w$ and $w^2$ respectively. As we progress one level higher up in the device hierarchy, we increment the exponent of the number $w$ by one. For a super logic region (SLR), which we find on devices with 2.5D packaging, we would have a weight of $w^3$. The number $w$ is an arbitrarily chosen number and is set to $5\ units$ in our experiments. For spatial distance threshold lower than $30\ units$, we obtained clusters which are too small to be of any significance. For edit distance thresholds larger than $10\ units$ we obtained poor groupings with the chosen number $w$. We set edit distance threshold to in the range from 1 to 10 units and spatial distance threshold to start from $30\ units$. In our experiments, a spatial distance threshold of $100\ units$ and edit distance threshold of $3\ units$ gave the best results.

\section{Experimentation}

\subsection{Experimental Setup}
We use the tool Hardware Analyzer (HAL) \cite{wallat2019highway} to perform necessary basic operations on the flattened netlist. HAL can accept a flattened netlist in Verilog for ASIC or FPGA \cite{wallat2019highway}. HAL's python interface and Python3 were used to implement the algorithm. NetworkX \cite{hagberg2008exploring} aided in performing graph operations and computing graph edit distances. Scikitlearn package \cite{scikit-learn} included functions to compute the NMI metric for checking cluster quality. Xilinx Vivado 2021.1 was used to synthesize the benchmarks to obtain a fully flattened hierarchy. The exported Verilog for the flattened hierarchy contains hierarchy information that will be used later as a reference hierarchy to compute the quality of clustering. 

The benchmarks chosen to test the algorithm are from a wide variety of netlists found on the OpenCores org site \cite{opencores}. They belong to arithmetic units, processors, DSP units, communication units and encoders. The designs were synthesized into fully flattened netlists. During synthesis, we exclude any DSP blocks as clustering of DSP blocks is out of scope of this paper. We allow LUTs, latches, registers, RAM blocks, SRL and MUX elements.

\subsection{Analysis}
We run our algorithm on the flattened netlists and obtain the groupings. We evaluate the groupings using Normalized Mutual information score (NMI). Normalized Mutual Information is a widely used metric to analyze the quality of clustering. "The Mutual Information is a measure of the similarity between two labels of the same data. Let the total number of samples be $N$, $|U_i|$ be the number of the samples in cluster $U_i$ and $|V_j|$ be the number of the samples in cluster $V_j$. The Mutual Information (MI) and Normalized Mutual Information (NMI) between clusterings $U$ and $V$ are given as \cite{thomas2006elements, scikit-learn}: "

\begin{displaymath}
    MI(U,V)=\sum_{i=1}^{|U|} \sum_{j=1}^{|V|} \frac{|U_i\cap V_j|}{N}
                \log\frac{N|U_i \cap V_j|}{|U_i||V_j|}
\end{displaymath}

\begin{displaymath}
    NMI(U,V)=\frac{{MI}(U, V)}{{mean}(H(U), H(V))}
\end{displaymath}

Each group is assigned a group id and the elements that belong to the same group share the same group id. We call the actual hierarchy exhibited by the design in the source Verilog as the reference hierarchy. A one dimensional array of group indices is obtained from a reference hierarchy at the word level and another array obtained from the inferred hierarchy are fed into a function to compute NMI. The array indices map to the design element indices. The element indices are in the same order in both the arrays. A design element with index 'i' with group id 'gi' would create an entry at the 'i'th position in the array with value 'gi' (0 with element 0, 33 with element 33).  NMI of less than 0.5 would mean that the obtained clustering has little information about the actual hierarchy. An NMI of 0.5 or greater tells us that we have managed to infer the actual hierarchy by a significant degree. 

\begin{figure}
    \centering
    \includegraphics[width=\linewidth, height=0.22\textheight]{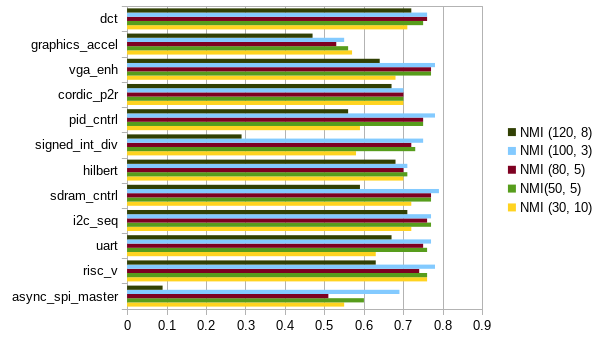}
    \caption{NMI computed for groupings with all types of elements included. The legend shows five sets of constraints. The first item in brackets is the average spatial distance threshold. The second item in the brackets is the average edit distance threshold. The NMI is highest  overall for the set with spatial distance threshold=100, edit distance threshold=3. Averages for NMI (30, 10), NMI (50, 5), NMI (80, 5), NMI (100, 3), NMI (120, 8) are 0.65, 0.71, 0.70, 0.73 and 0.56 respectively.}
    \label{nmi_all_chart}
\end{figure}

\begin{figure}
    \centering
    \includegraphics[width=\linewidth, height=0.24\textheight]{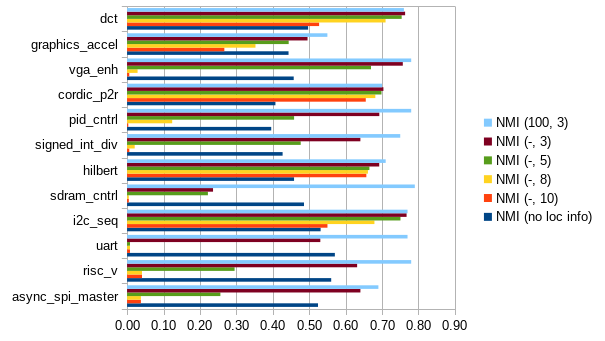}
    \caption{The NMI computed for the case where we do not use location information NMI (no loc info) is lower than the NMI computed when location information is used. The '-' in the brackets indicate location information is not used in the second stage of the algorithm. The average NMI computed for when proximity of sites is considered is still the highest among all. The average NMI computed for NMI (no loc info), NMI (-, 10), NMI (-, 8), NMI (-, 5), NMI (-, 3) and NMI (100, 3) are 0.48, 0.23, 0.28, 0.47, 0.63 and 0.74 respectively.}
    \label{nmi_no_loc_vs_loc_chart}
\end{figure}


\begin{figure*}[!h]
    \centering
    \includegraphics[width=0.85\linewidth]{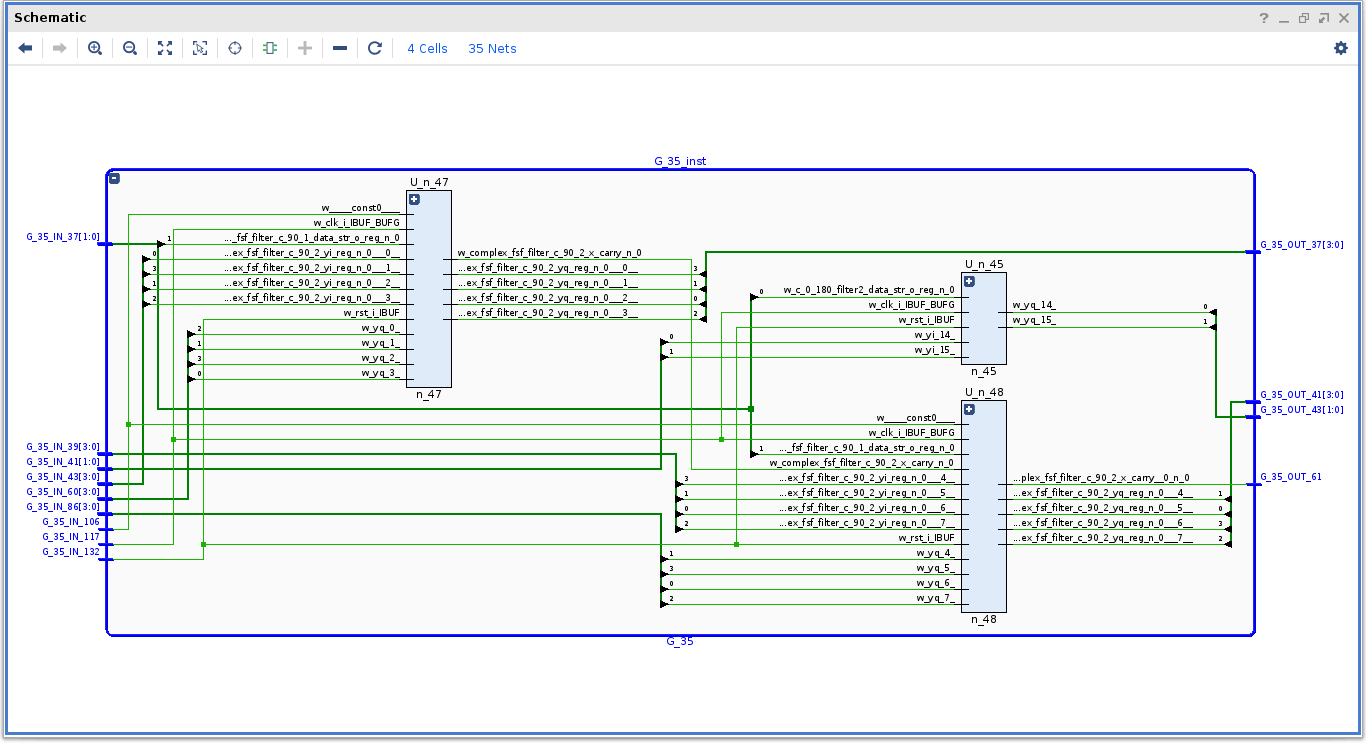}
    \caption{A grouping from a hilbert transform unit with (spatial distance threshold=50, edit distance threshold=5). N\_47 has indices 0, 1, 2, 3 of output in G\_35\_OUT\_37 and N\_48 has indices 4, 5, 6, 7 of the output in G\_35\_OUT\_61. The input w\_yq is fed to the two modules. Indices 0, 1, 2, 3 of w\_yq go to N\_47 and indices 4, 5, 6, 7 of w\_yq go to N\_48. N\_45 and N\_48 share inputs and output carry net from N\_47 is fed into N\_48.}
    \label{hilbert_g35}
\end{figure*}

In addition to grouping the design elements we have also grouped the nets together. Each net has a source design element. We record the source element's type, Boolean equation and the location information. We check if two nets have the same source element's type and Boolean equation. Two nets satisfying these criteria could belong to different modules performing the same operation (for example, consider two different instances of an adder module). We also check if the two source elements driving these two nets are in close proximity on the device. If the nets satisfy these criteria, the nets belong to the same word. Otherwise the two nets belong to different words and in different modules. We present few sample groupings to provide a qualitative measure.

\subsection{Results and Discussion}

The charts show that when average spatial distance threshold is set to 100 and average edit distance threshold is set to 3, we get the maximum overall NMI for groupings including all types of elements and for register groupings. We also observe that for smaller magnitudes of thresholds on spatial distance, we could allow larger thresholds on edit distance and still obtain an acceptable NMI. As we move far away from a location we could begin observing the same structures which belong to a different module. Two structures in close proximity are distinguished by the edit distance. We expect similar structures to be nearby as these structures would operate on a word. A dissimilarity in the structure would indicate a different operation.

\begin{table}
    \centering
    \caption{Evaluation of cluster quality (average spatial distance threshold=100, average edit distance threshold=3)}
    \resizebox{\linewidth}{!}{
    \begin{tabular}{|l|r|r|r|r|r|r|r|}
    \hline
        Design & Cell count & NMI & TP & TN & FP & FN & Accuracy  \\ \hline
        async\_spi\_master & 153 & 0.69 & 214 & 21,280 & 1486 & 276 & 0.92  \\ 
        risc\_v & 273 & 0.78 & 790 & 71,072 & 1646 & 748 & 0.97  \\ 
        uart & 291 & 0.77 & 852 & 80,594 & 1366 & 1578 & 0.97  \\ 
        i2c\_seq & 436 & 0.77 & 790 & 184,128 & 2168 & 2574 & 0.97  \\ 
        sdram\_cntrl & 562 & 0.79 & 2486 & 305,254 & 3270 & 4272 & 0.98  \\ 
        fpadd & 741 & 0.63 & 1932 & 508,452 & 1722 & 36,234 & 0.93  \\ 
        hilbert & 1044 & 0.71 & 2896 & 1,065,780 & 10,060 & 10,156 & 0.98  \\ 
        signed\_int\_div & 1364 & 0.75 & 2948 & 1,835,238 & 6980 & 13,966 & 0.99  \\ 
        pid\_cntrl & 1428 & 0.78 & 2152 & 2,018,648 & 5554 & 11,402 & 0.99 \\ 
        cordic\_p2r & 1655 & 0.70 & 5060 & 2,680,072 & 11,048 & 41,190 & 0.98  \\ 
        vga\_enh & 2155 & 0.78 & 7550 & 4,575,814 & 40054 & 18452 & 0.99 \\ 
        graphics\_accel	& 3056 & 0.55 & 40,868 & 8,539,754	& 415,878 & 339,580 & 0.92 \\ 
        dct	& 3607 & 0.76	& 7148 & 12,869,184 & 19,550 & 110,960 & 0.99 \\ \hline
        Average & ~ & 0.73 & ~ & ~ & ~ & ~ & 0.97  \\ \hline
    \end{tabular}}
    \label{designs_all_measures}
\end{table}

\begin{table*}
    \centering
    \caption{Evaluation of cluster quality for design dbl\_clk\_ifft when timing constraints and synthesis directives are applied (average spatial distance threshold=100, average edit distance threshold=3)}
    \resizebox{0.8\linewidth}{!}{
    \begin{tabular}{|l|r|r|r|r|r|r|r|r|r|}
    \hline
        Clock period/synthesis directive & Cell count & NMI & TP & TN & FP & FN & Accuracy \\ \hline
        12 ns & 13,080 & 0.818 & 40,160 & 170,708,002 & 93,848 & 231,310 & 0.998  \\ 
        11 ns & 13,080 & 0.818 & 39,976 & 170,707,178 & 94,672 & 231,494 & 0.998  \\ 
        10 ns & 13,180 & 0.815 & 37,844 & 173,342,756 & 91,646 & 226,974 & 0.998  \\ 
        9 ns & 13,180 & 0.815 & 37,902 & 173,342,764 & 91,638 & 226,916 & 0.998  \\ 
        8 ns & 13,180 & 0.815 & 37,816 & 173,341,632 & 92,770 & 227,002 & 0.998  \\ 
        6 ns & 13,180 & 0.814 & 37,536 & 173,343,480 & 90,922 & 227,282 & 0.998  \\ 
        5 ns & 13,180 & 0.815 & 37,884 & 173,343,310 & 91,092 & 226,934 & 0.998  \\ 
        10\_ns\_area\_optimized\_high & 23,960 & 0.772 & 58,446 & 572,658,232 & 191,992 & 1,148,970 & 0.998  \\ 
        10\_ns\_alternate\_routing & 13,747 & 0.805 & 36,904 & 188,585,126 & 98,754 & 245,478 & 0.998  \\ 
        10\_ns\_fewer\_carry\_chains & 13,088 & 0.819 & 39,870 & 170,920,384 & 90,796 & 231,606 & 0.998  \\ 
        10\_ns\_logic\_compaction & 13,080 & 0.818 & 40,504 & 170,705,194 & 96,656 & 230,966 & 0.998  \\ 
        10\_ns\_performance\_optimized\_high & 13,411 & 0.812 & 38,576 & 179,447,488 & 88,816 & 266,630 & 0.998  \\ \hline
        Average & ~ & 0.811 & ~ & ~ & ~ & ~ & 0.998  \\ \hline
    \end{tabular}}
    \label{designs_constraints}
\end{table*}

The NMI was computed for all the benchmarks and the largest NMI obtained turned out to be 0.73 for groupings with all element types included as can be seen in Figure \ref{nmi_all_chart}. Figure \ref{nmi_no_loc_vs_loc_chart} shows the relative differences in the quality of grouping based on how much location information is used. We run the greedy modularity graph clustering algorithm  which uses no location data on the elements. We also run our algorithm without proximity information utilized after site level groupings. We compare these groupings with the grouping obtained for NMI (100, 3) which is the best constraint found so far. The chart shows that location information when considered indeed improves the NMI. Table \ref{designs_all_measures} shows the pairwise true positives (TP), true negatives (TN), false positives (FP) and false negatives (FN) and accuracy measures for all designs with the average spatial distance threshold and average edit distance threshold set as (100, 3) which gave the highest NMI. "A true positive (TP) decision assigns two similar samples to the same cluster, a true negative (TN) decision assigns two dissimilar samples to different clusters. A (FP) decision assigns two dissimilar samples to the same cluster. A (FN) decision assigns two similar samples to different clusters. The Rand index or accuracy measures the percentage of decisions that are correct \cite{cluster_quality}." NMI is the more reliable measure \cite{cluster_quality}.

\[ Accuracy = \frac{TP + TN}{TP + TN + FP + FN}\]

A design dbl\_clk\_ifft, was synthesized with different timing constraints and synthesis directives on the same device xc7a12tcsg325-1. Location information was used in grouping elements in all cases with average spatial distance threshold set to 100 and average edit distance threshold set to 3. There was no significant impact on NMI and accuracy due to timing constraints or synthesis directives as shown in Table \ref{designs_constraints}.


Figure \ref{hilbert_g35} shows a grouping from a hilbert transform unit with (spatial distance threshold=50, edit distance threshold=5). N\_48 gets inputs with indices 4, 5, 6 and 7 from a word named w\_complex\_fsf\_filter\_c\_90\_2\_yi\_reg\_n\_0\_. N\_47 gets inputs with indices 0, 1, 2, and 3 from the same word. We observe a similar case with input word w\_yq. Four bits of a word are fed to one submodule N\_47 and the other four are fed to another submodule N\_48. The outputs from these two submodules will belong to the same word and these submodules are correctly grouped together.
\section{Conclusion}
We have presented an FPGA design element grouping algorithm which requires no prior knowledge about the circuit. The algorithm is not specific to IP block identification or exclusive to a certain design element type. The algorithm uses location information of the design elements utilized in an FPGA design and the graph edit distance measure together to help infer the hierarchy of the design. The NMI is improved when the constraint on spatial distance is relaxed and when the constraint on graph edit distance is tightened. The algorithm uses the same set of constraints for all the designs. Our experiments show us that for set of constraints (spatial distance threshold=100, edit distance threshold=3), we get maximum average NMI for groupings with all elements included and for register groupings. Our experiments show us that the location information does contain considerable amount of information about the higher level structures present in the design. Location information alone is not sufficient to work with and another source of information such as the graph edit distance is used to group elements. 

The clustering of elements or partitioning the netlist helps in dividing a large and complex problem of specification discovery into simpler subproblems. After grouping elements, we need to identify the functionality of the grouping, where the cluster may belong to an operator or a state machine. Subramanyan et al. \cite{subramanyan2013reverse} propose structural and functional methods to identify functionality in an ASIC netlist. We focus on identifying operators and state machines in our future work. 

\section*{ACKNOWLEDGEMENT}
This work was supported in part by the National Science Foundation under IUCRC-1916762, and CHEST (Center for Hardware Embedded System Security and Trust) industry funding.

\nocite{*}
\bibliographystyle{ieeetr}

\end{document}